\documentclass[APA,alpha-refs]{wiley-article}

\usepackage{siunitx}

\papertype{Original Article}
\paperfield{Journal Section}

\title{Generative AI alone may not be enough: Evaluating AI Support for Learning Mathematical Proof}

\author[1]{Eason Chen\authfn{1}}
\author[1]{Sophia Judicke}
\author[1]{Kayla Beigh}
\author[1]{Xinyi Tang}
\author[1]{Zimo Xiao}
\author[1]{Chuangji Li}
\author[1]{Shizhuo Li}
\author[1]{Reed Luttmer}
\author[1]{Shreya Singh}
\author[1]{Maria Yampolsky}
\author[1]{Naman Parikh}
\author[1]{Yi Zhao}
\author[1]{Meiyi Chen}
\author[1]{Scarlett Huang}
\author[1]{Anishka Mohanty}
\author[1]{Gregory Johnson}
\author[1]{John Mackey}
\author[1]{Jionghao Lin}
\author[1]{Ken Koedinger}

\affil[1]{Carnegie Mellon University, Pittsburgh, PA, 15213, United States}

\corraddress{Eason Chen, Human-Computer Interaction Institute, Carnegie Mellon University, Pittsburgh, PA, 15213, United States}
\corremail{EasonC13@cmu.edu}

\runningauthor{Chen et al.}

\begin{document}

\begin{frontmatter}
\maketitle


\begin{abstract}


We evaluate the effectiveness of LLM-Tutor, a large language model (LLM)-powered tutoring system that combines an AI-based proof-review tutor for real-time feedback on proof-writing and a chatbot for mathematics-related queries. Our experiment, involving 148 students, demonstrated that the use of LLM-Tutor significantly improved homework performance compared to a control group without access to the system. However, its impact on exam performance and time spent on tasks was found to be insignificant. Mediation analysis revealed that students with lower self-efficacy tended to use the chatbot more frequently, which partially contributed to lower midterm scores. Furthermore, students with lower self-efficacy were more likely to engage frequently with the proof-review-AI-tutor, a usage pattern that positively contributed to higher final exam scores. Interviews with 19 students highlighted the accessibility of LLM-Tutor and its effectiveness in addressing learning needs, while also revealing limitations and concerns regarding potential over-reliance on the tool. Our results suggest that generative AI alone like chatbot may not suffice for comprehensive learning support, underscoring the need for iterative design improvements with learning sciences principles with generative AI educational tools like LLM-Tutor.


\end{abstract}
\end{frontmatter}

\section{Introduction}



Mathematical proficiency is important in higher education, particularly in STEM fields \citep{golding2018mathematics, whitney2022role, maass2019role}. A critical component of demonstrating mathematical proficiency is proof-writing, which fosters logical reasoning and a deep conceptual understanding beyond rote memorization \citep{thurston1995proof, schoenfeld2010series, stewart2019student}. Proof-writing is a common assessment method in undergraduate entry-level math courses but is often a significant hurdle for many students. Students frequently struggle with the proof-writing process and receive limited personalized support, making it difficult to overcome their challenges effectively \citep{stewart2019student}. These difficulties can lead to decreased confidence, disengagement, and gaps in mathematical understanding. Therefore, these challenges underscore the urgent need for innovative approaches to support and enhance proof-writing skills in mathematics education.

Large Language Models (LLMs) have opened up new possibilities for creating personalized, scalable, and interactive learning systems in education \citep{wardat2023chatgpt, frieder2024mathematical, li2024bringing}. By providing real-time feedback and adaptive support, these systems have the potential to transform how students approach and engage with complex subjects \citep{stamper2024enhancing, dai2024assessing, fujita2018learners}. However, more evidence is needed to evaluate the effectiveness of LLM-powered learning systems in real-world educational settings \citep{chen2024systematic, li2024bringing}. Questions remain regarding whether these systems can consistently and meaningfully enhance learning outcomes.

Motivated by the current questions, we developed LLM-Tutor, an LLM-powered tutoring system designed to assist students in a discrete mathematics course. LLM-Tutor provides real-time feedback on proof-writing tasks to improve students’ performance, self-efficacy, and time management. This study presents an empirical evaluation of the impact of LLM-Tutor, combining both quantitative and qualitative methods to gain a comprehensive understanding of its effects.

To guide our investigation, we pose the following \textbf{R}esearch \textbf{Q}uestions:

\begin{itemize}
    \item \textbf{RQ1}: How does the use of LLM-Tutor affect students' performance in a mathematics course, as measured by exam and homework scores?
    \item \textbf{RQ2}: How does LLM-Tutor influence students' perceived time spent in the coursework?
     \item \textbf{RQ3}: Does the use of LLM-Tutor mediate the relationship between students' self-efficacy and their exam score?
    \item \textbf{RQ4}: How do students perceive the advantages and limitations of LLM-Tutor?
\end{itemize}

\section{Related Works}

\subsection{Large Language Models in Education}

Large Language Models (LLMs) are advanced AI systems designed to process and generate human-like text by training on vast datasets of written content \citep{kamath2024large}. These models are defined by their extensive number of parameters—often in the billions—which enable them to accurately capture and replicate complex linguistic patterns \citep{openai2023gpt4}. LLMs work by producing outputs in response to text prompts, which guide their response to specific tasks or queries \citep{chen2024systematic}.

Prompt techniques such as Few-Shot Learning \citep{brown2020language} and Chain-of-Thought (CoT) prompting \citep{wei2022chain} have significantly extended the capabilities of LLMs \citep{chen2024systematic}. Few-shot learning allows models to generalize patterns from a limited number of examples, while CoT prompting supports step-by-step reasoning to improve logical coherence and accuracy. OpenAI’s latest o1 model exemplifies these advances by incorporating automatic chain-of-thought reasoning. This model is claimed to have PhD-level reasoning ability and has achieved remarkable results in mathematical problem solving, including a success rate 83\% in the International Mathematics Olympiad, compared to the GPT-4 13\% \citep{openai2024o1}.

Existing research on LLM in education has shown mixed results. Although some studies report increased self-efficacy and engagement \citep{wardat2023chatgpt, shahzad2024exploring}, others raise concerns about overreliance and potential negative impacts on learning outcomes \citep{zhai2024effects}. Moreover, most existing research focuses on short-term qualitative feedback or controlled experimental settings, leaving questions about their real-world applicability and long-term effects on learning outcomes. For example, \citet{shahzad2024exploring} found through qualitative analysis that LLMs significantly enhance student performance in higher education, improving self-efficacy, fairness, ethics, and creativity. Similarly, \citet{wu2024promoting} showed that LLM-based learning tools support self-regulation and knowledge construction in mathematics classes over a 10-day period.

This study seeks to address these gaps by providing an evidence-based analysis of the effectiveness of LLM-powered tutoring systems in real-world educational settings.

\subsection{Intelligent Tutoring Systems for Proof Writing}

Intelligent Tutoring Systems (ITS) are computer-based learning environments that employ artificial intelligence techniques to provide learners with adaptive feedback \citep{anderson1985intelligent}. ITS can provide personalized instruction by leveraging predefined rules and have proven effective in well-structured domains like geometry \citep{aleven2002effective, stylianides2024proof, fujita2018learners}. However, extending such systems to more complex, open-ended proofs, such as those in discrete mathematics, requires extensive domain-specific logic, making development challenging \citep{fujita2018learners}.

In contrast, Large Language Models (LLMs) can adapt dynamically to diverse contexts without relying on fixed rule sets \citep{brown2020language}. Their flexible feedback capabilities have recently enabled real-time, personalized tutoring at scale, as seen in systems like Khanmigo \citep{khan2024khanmigo, shetye2024evaluation}. Yet, the effectiveness of these kind of LLM powered tutoring system in guiding advanced proof writing remains largely unexplored.

Our study investigates LLM-Tutor, an LLM-powered tutoring system designed to support complex proof tasks in a real-world discrete mathematics course. By examining its performance, we aim to understand how LLMs can complement traditional ITS and better meet the challenges posed by open-ended mathematical reasoning.

\subsection{Self-Efficacy and Its Role in Learning}

Self-efficacy, defined as an individual's belief in their capability to succeed in specific tasks, is a well-established predictor of academic performance and engagement \citep{bandura1997self}. It influences how students approach challenges, the effort they expend, and their persistence in the face of difficulties \citep{zimmerman2000self, pintrich1990motivational}. In mathematical learning contexts, self-efficacy is particularly critical, as it affects students' willingness to engage with complex problem-solving tasks such as proof writing \citep{hackett1985role}.

Research has shown that students with higher self-efficacy are more likely to employ effective learning strategies and seek out resources for support \citep{schunk1991self}. Conversely, those with low self-efficacy may avoid challenging tasks or over-rely on external support, potentially hindering their independent learning \citep{zimmerman2000self, zakariya2022improving}. This dynamic makes self-efficacy a key factor to consider when evaluating the effectiveness of AI-powered educational tools, where students with different self-efficacy might adopt various strategies in using it, leading to different impacts on their learning performance.

\section{System Developments}

We developed LLM-Tutor to address the above-mentioned challenges and provide students with high-quality, timely feedback on open-ended, proof-writing math assignments and answer math-related questions. LLM-Tutor features two key functionalities: Proof-Review-AI-Tutor providing feedback on students' written proofs, and the Chatbot capable of responding to student questions in real time. This section introduces these two functionalities in detail.

\subsection{Proof-Review-AI-Tutor to Guide Students' Open-Ended Proofs Writing}

When providing feedback on students' writing, it is crucial to ensure that the feedback is explainable. Drawing on previous research that underscores the value of visual cues (e.g., highlighting) in improving the explainability of AI feedback \citep{vasconcelos2023explanations}, we integrated visual highlights into our design. As shown in Figure 1, AI feedback not only offers overall comments on the student's proof (e.g., \textit{``You are on the right track, but it seems you misunderstand the concept of XXX''}), but also highlights specific problematic sections in the proof (e.g., "This equation does not make sense"). This approach enables students to easily identify areas for improvement in their proofs.

\begin{figure}
    \centering
    \includegraphics[width=1\linewidth]{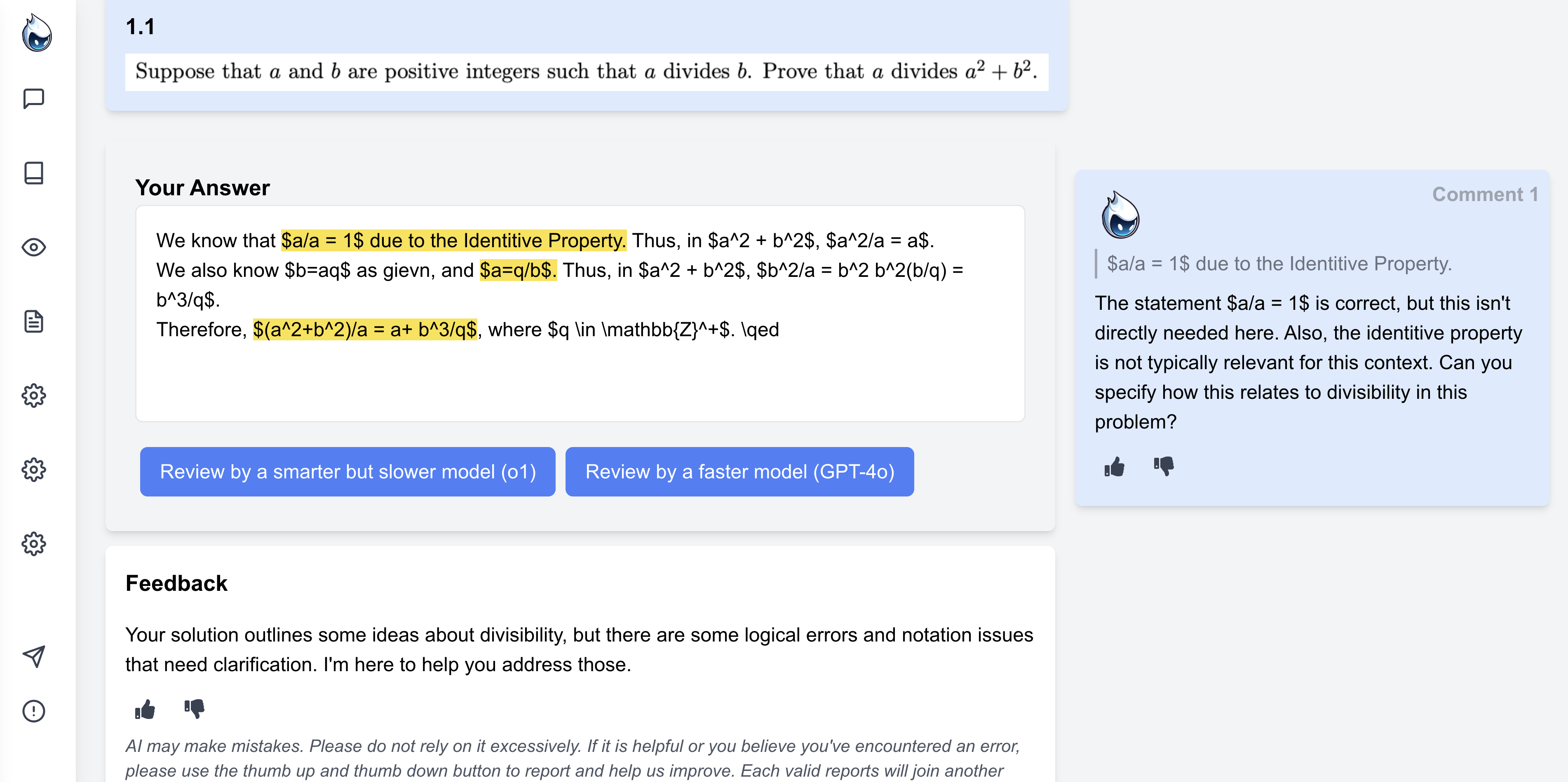}
    \caption{The interface of Proof-Review-AI-Tutor where it reviews student homework proof assignments.}
    \label{fig:homework_reviewer}
\end{figure}

In the Proof-Review-AI-Tutor, the correct answers for each question are input into the system as prompts for in-context learning. To prevent students from getting answers, we only support formatted output and do not allow open-ended conversations to prevent students from attempting to extract answers. Before the assigned coursework is released, the collaborating course instructors provide us with the assignments and their correct answers in LaTeX format. We then update the system accordingly for student use.

When using the Proof-Review-AI-Tutor, students can freely choose between using OpenAI's O1, which is more powerful but slower, or GPT-4o, which is less powerful but provides more immediate feedback. 

With a designated prompt, LLM can act as a professional math tutor, focusing on the mathematical logic issues in the student's response rather than picking on minor errors. For example, in Figure \ref{fig:homework_reviewer}, LLM ignored the typo "gievn" and concentrated on the overall correctness of the student's answer. Because of the length constraints, the prompt we used for Proof-Review-AI-Tutor can be found in Appendix A.

Subsequently, on the front end, we process the outputs from LLM in JSON format and display them as highlighted sections. When a user clicks on a highlighted part, a specific feedback is displayed for that part, as shown in Figure \ref{fig:homework_reviewer}.

\subsection{AI Chatbot for Answering student's question}

Given the popularity of chatbot-based interactions for LLMs like ChatGPT, LLM-Tutor also includes a built-in chatbot feature with GPT-4o that allows students to engage in conversations. We designed the chatbot interface to closely resemble ChatGPT’s interface, making it intuitive for students to use and enabling us to track their interactions. 

This chatbot does not have access to course content; instead, we use a simple prompt to instruct the chatbot to focus on math-related discussions. During the preliminary testing phase, we found that the chatbot often provided answers directly for math questions. To address this, we specify in the prompt that the chatbot does not directly provide answers or proof steps. Instead, it was instructed to use the Socratic questioning method or scaffolding questions to guide the students in discovering the answers themselves. Because of the length constraints, the prompt we used for Chatbot can be found in Appendix B.

\begin{figure}
    \centering
    \includegraphics[width=1\linewidth]{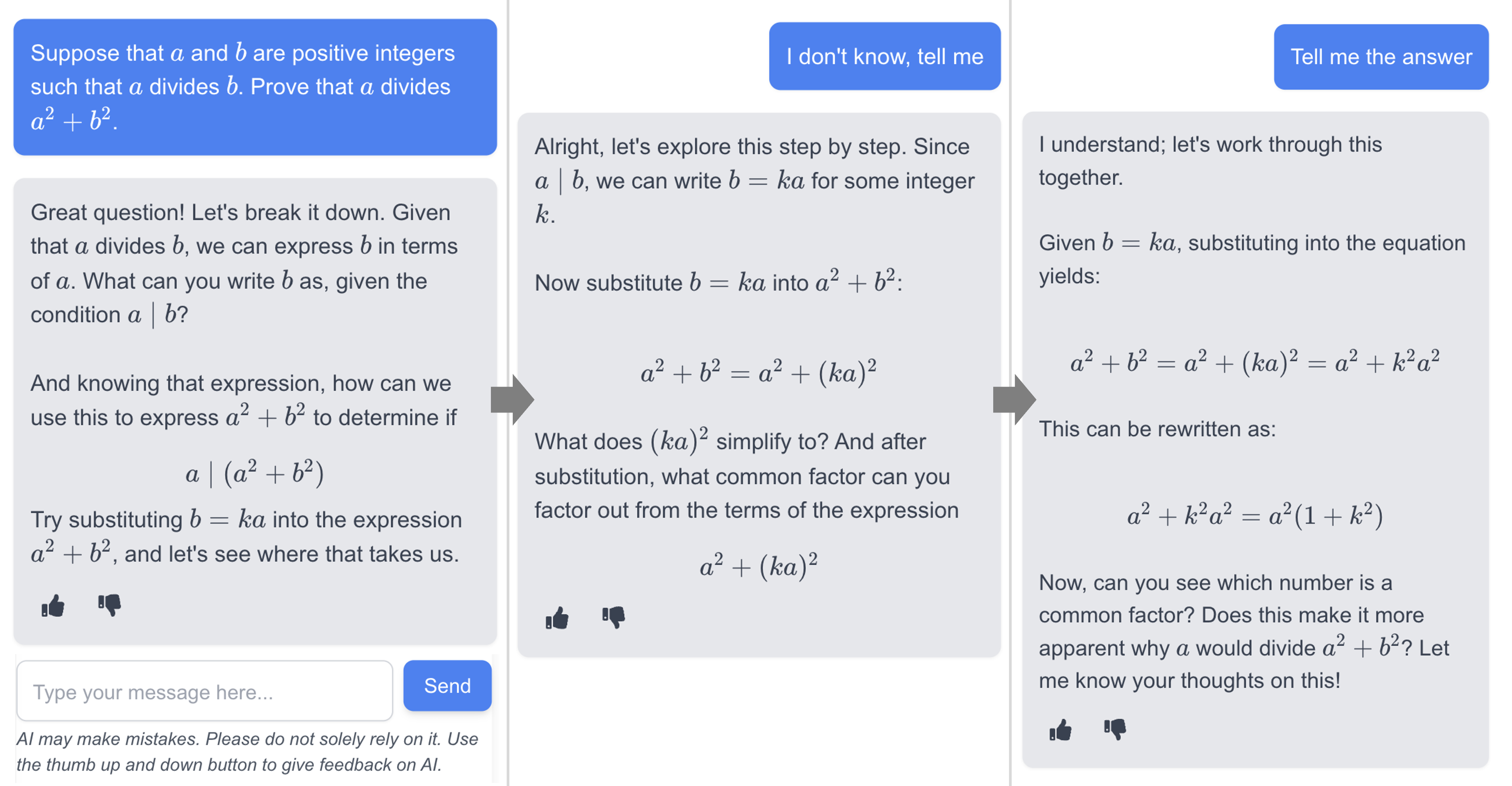}
    \caption{Screenshot of the LLM-Tutor chatbot's conversation interface. The message shows that the chatbot avoids directly providing answers to students but will give them upon request specifically. Due to length limitations, the chat history is displayed vertically from top to bottom and arranged from left to right.}
    \label{fig:LLM-Tutor_chat}
\end{figure}

Compared to directly using ChatGPT, Claude, or Genimi, our prompt prevents students from receiving direct answers when they paste their homework questions. However, we are not outright prohibiting students from getting answers. If a student persists in asking, such as repeatedly saying, "Give me the answer" after entering their question, they can still obtain the solution or the steps of the proof. An example of this conversation is shown in Figure \ref{fig:LLM-Tutor_chat}.

\section{Research Methods}

\subsection{Participant Recruitment and Research Design}
Our research was approved by the Institutional Review Board (IRB). We collaborated with a research university located on the East Coast of the United States to deploy LLM-Tutor in their classrooms. The class involved an intro-level discrete math course with 308 students. Since assignments for this course were submitted using LaTeX, students are encouraged to paste their LaTeX homework submission into the Proof-Review-AI-Tutor to seek feedback before submission and discuss math-related topics with the chatbot.

To recruit participants, we made announcements in class and through the course mailing list, inviting students to participate in the study. Interested students were required to complete a Google form in which they reported their current learning status, strategies, and the time they spent on the course. The students also provided their consent for LLM-Tutor to use their data for research purposes.

The participants were then randomly divided into the experimental group and the control group. The experimental group gained access to LLM-Tutor immediately after the first midterm exam and homework 4, while the control group was granted access only after the second midterm exam.

We recruited a total of 155 students, with 78 assigned to the control group and 77 to the experimental group. After the midterm withdrawal period, the experimental group decreased to 73 students, while the control group had 75 remaining students. By the time of the final exam, the number of students in the experimental group remained unchanged, but the control group decreased to 72 students.

\subsection{Data Collection}

We collect student grades across 10 homeworks, 3 midterm exams, and the final exam score.

Then, in LLM-Tutor, we collect interaction logs between students and the Chatbot, as well as the Proof-Review-AI-Tutor, and analyze them based on usage frequency. We categorized the frequency of usage according to the intervals between exams. For example, to analyze Midterm 3, we used the interaction logs recorded between Midterm 2 and Midterm 3. Similarly, for the final exam, we used the logs recorded between Midterm 3 and the final exam.

When students joined after the first midterm, they must complete a survey using the same Google Form before submitting their consent form. The survey collects self-reported data on the amount of time they spend on the course each week and their self-efficacy in the course. The self-efficacy questions are adapted from \cite{pintrich1990motivational}, with a Cronbach's alpha of 0.91 in the data we collected. After the second midterm exam, we invite students to complete a follow-up survey to report their time spent on the course and their SUS score with LLM-Tutor.


We then conducted semi-structured interviews with 19 students to gain deeper insights into their experiences using LLM-Tutor. The interviews focused on how LLM-Tutor influenced their learning habits, its comparision on traditional teaching resources such as teaching assistants and peers, and its effects on their understanding of course material and self-efficacy. In addition, we explore the role of the tool in shaping students' motivation and engagement with the course and discuss LLM-Tutor's limitations.






\subsection{Data Analysis}
To analyze our data, we conducted several tests to explore different aspects of the study with SPSS.

First, we used t-tests to examine differences in exam performance and homework performance between the control and experimental groups. Next, we performed a correlation analysis to explore relationships between exam grades, usage of Proof-Review-AI-Tutors, usage of chatbots, and the number of response requests. A second correlation analysis was performed to examine the connections between these variables and self-efficacy as well as homework grades.

Furthermore, we applied multiple regression to predict student scores in Midterm 3, using Proof-Review-AI-Tutor usage, chatbot usage, and answer request quantity as predictors. Finally, we performed a mediation analysis to investigate the relationships between chatbot usage, self-efficacy, response request quantity, and midterm scores.

Note that for correlation, regression, and mediation analysis, we focus on the interval between midterm 2 and 3 when both the control group and experiment group have access to LLM-Tutor, to maximize the sample size.

\section{Results}

\subsection{Compare homework grade across student groups}

\begin{figure}[h]
    \centering
    \includegraphics[width=1\linewidth]{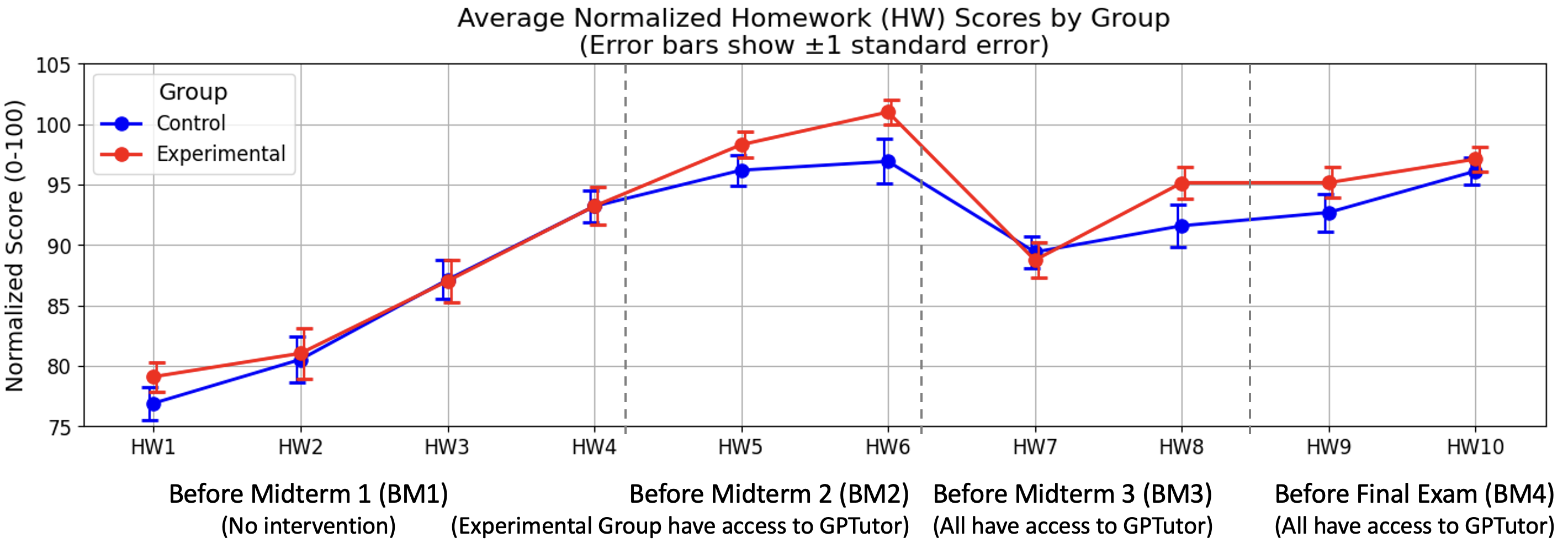}
    \caption{Normalized Homework scores of students between different groups in Class A. The results show that during the Before Midterm 2 period, when the experimental group had access to LLM-Tutor but the control group did not, the experimental group performed significantly better on homework on average than the control group.}
    \label{fig:jmackey_homework_grade}
\end{figure}

Students' homework (HW) grades, normalized as percentages, are presented in Figure \ref{fig:jmackey_homework_grade}. Prior to the use of LLM-Tutor (HW1, HW2, HW3, HW4), the experimental group showed no significant difference in scores compared to the control group (t(146) = 0.64, p = .524). However, after LLM-Tutor was introduced, the experimental group's average scores (M = 99.64, SD = 8.03) for HW5 and HW6 significantly exceeded those of the control group (M = 96.55, SD = 11.66) (t(146) = 2.19, p = .030). Once the control group also began using LLM-Tutor, the homework performance of both groups became comparable, with no significant differences observed in the average score before midterm 3 (t(146) = 1.25, p = .212) or before final (t(144) = 1.27, p = .206). These findings indicate that the use of LLM-Tutor may enhance students' performance on homework tasks.

\begin{figure}[h!]
    \centering
    \includegraphics[width=1\linewidth]{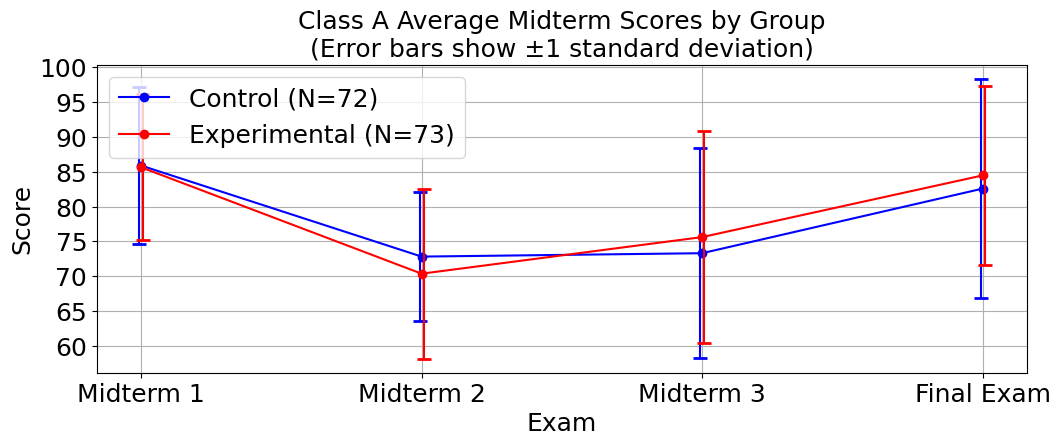}
    \caption{Midterm scores of students between different groups in Class A. No condition-based significant differences were observed either prior to the LLM-Tutor implementation (Midterm 1) or after (Midterms 2 and 3).}
    \label{fig:midterm}
\end{figure}

\subsection{Compare exam grades across student groups}

Midterm exam scores for the control and experimental groups are shown in Figure \ref{fig:midterm}. An independent samples t-test was conducted to compare the performance of the experimental group (students who began using LLM-Tutor immediately after Midterm 1) and the control group (students who did not use LLM-Tutor until after Midterm 2 had concluded) across the three midterm exams and the final exam. No statistically significant differences were found, indicating that the use of LLM-Tutor did not significantly impact students' performance on the exams.

\subsection{Correlation of Frequency, Self Efficacy, and Homework and Exam Grades}

\begin{figure}[h!]
    \centering
    \includegraphics[width=1\linewidth]{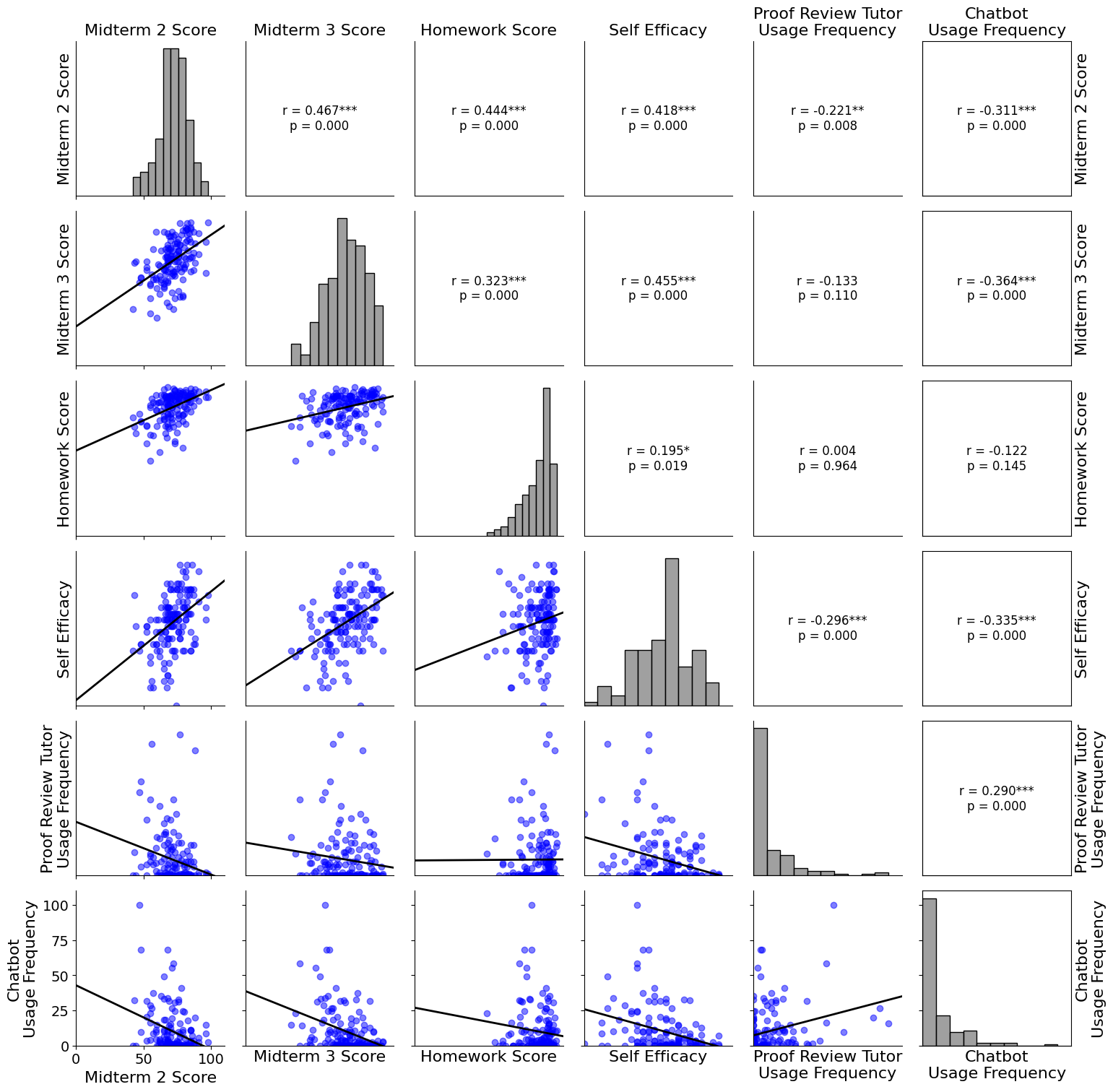}
    \caption{Correlations Between Academic Performance (Midterms, Homework), LLM-Tutor Usage (Proof-Review-AI-Tutor \& Chatbot), and Self-Efficacy, as well as their distribution. The Proof-Review-AI-Tutor Usage Frequency, and Chatbot Usage Frequency was the frequency between the interval of Midterm 2 and Midterm 3. The Homework was the sum of Homework 6 and Homework 7. Note that the distribution for Self Efficacy, Proof-Review-AI-Tutor Usage Frequency, and Chatbot Usage Frequency is normalized to the range of 0 \- 100.}
    \label{fig:correlation}
\end{figure}

We conducted a deeper analysis of the correlation between students' usage frequency of the Proof-Review-AI-Tutor and chatbot features on the LLM-Tutor platform, including their performance on exams and homework assignments, as well as their Self-efficacy. The correlation results for data that combined both control and experimental groups are presented in Figure \ref{fig:correlation}. We also analyzed the data for the control and experimental groups separately, and the significance of the results was the same as that of the combined analysis. To save space, we present only the data from after the second midterm exam to before the third midterm exam. During this period, both the experimental and control groups had access to LLM-Tutor.

The correlation results indicate that students' homework grades show little to no correlation with their frequency of using various LLM-Tutor features. Additionally, prior midterm 2 exam scores and self-efficacy were negatively correlated with the frequency of LLM-Tutor usage. However, during Midterm 3, the frequency of using the Proof-Review-AI-Tutor was no longer significantly negatively correlated to students' exam scores.

\subsection{Multiple Regression on Usage Frequency and Academic Performance}

\begin{table}[h!]
\centering
\caption{Regression Analysis Results for Predicting Midterm 3 Score}
\label{tab:regression_midterm3}
\begin{tabular}{lcccccc}
\hline
\textbf{Model} & \textbf{Variable} & \textbf{B} & \textbf{Std. Error} & \textbf{Standardized Beta} & \textbf{t} & \textbf{p-value} \\
\hline
1 & (Constant) & 78.179 & 1.493 & - & 52.349 & .000 \\
  & Chatbot Usage Frequency & -0.056 & 0.013 & -0.356 & -4.358 & .000 \\
  & Proof-Review-AI-Tutor Usage Frequency & -0.027 & 0.074 & -0.030 & -0.369 & .713 \\
\multicolumn{7}{l}{\textbf{Notes for Model 1:} $R = .366$, $R^2 = .134$, Adjusted $R^2 = .121$, Std. Error of the Estimate = 14.1474} \\
\hline
2 & (Constant) & 52.701 & 5.361 & - & 9.831 & .000 \\
  & Chatbot Usage Frequency & -0.040 & 0.012 & -0.250 & -3.184 & .002 \\
  & Proof-Review-AI-Tutor Usage Frequency & 0.049 & 0.070 & 0.054 & 0.698 & .486 \\
  & Self Efficacy & 5.088 & 1.034 & 0.388 & 4.919 & .000 \\
\multicolumn{7}{l}{\textbf{Notes for Model 2:} $R = .510$, $R^2 = .261$, Adjusted $R^2 = .245$, Std. Error of the Estimate = 13.1164} \\
\hline
3 & (Constant) & 25.524 & 10.043 & - & 2.542 & .012 \\
  & Chatbot Usage Frequency & -0.037 & 0.012 & -0.232 & -3.032 & .003 \\
  & Proof-Review-AI-Tutor Usage Frequency & 0.032 & 0.068 & 0.035 & 0.462 & .645 \\
  & Self Efficacy & 4.513 & 1.019 & 0.344 & 4.428 & .000 \\
  & Homework Average Score & 0.327 & 0.103 & 0.228 & 3.163 & .002 \\
\multicolumn{7}{l}{\textbf{Notes for Model 3:} $R = .557$, $R^2 = .310$, Adjusted $R^2 = .290$, Std. Error of the Estimate = 12.7167} \\
\hline
4 & (Constant) & 11.108 & 11.196 & - & 0.992 & .323 \\
  & Chatbot Usage Frequency & -0.030 & 0.012 & -0.189 & -2.494 & .014 \\
  & Proof-Review-AI-Tutor Usage Frequency & 0.048 & 0.067 & 0.053 & 0.708 & .480 \\
  & Self Efficacy & 3.207 & 1.100 & 0.244 & 2.915 & .004 \\
  & Homework Average Score & 0.179 & 0.112 & 0.124 & 1.604 & .111 \\
  & Midterm 1 & 0.150 & 0.120 & 0.108 & 1.243 & .216 \\
  & Midterm 2 & 0.288 & 0.124 & 0.207 & 2.325 & .022 \\
\multicolumn{7}{l}{\textbf{Notes for Model 4:} $R = .596$, $R^2 = .355$, Adjusted $R^2 = .327$, Std. Error of the Estimate = 12.3809} \\
\hline
\end{tabular}
\end{table}

To investigate the relationship between students' usage frequency of LLM-Tutor features, their midterm and homework performance, and their self-efficacy, we conducted multiple regression analyses. We aimed to examine how different predictors influenced students' \textbf{Midterm 3 scores} and whether self-efficacy and prior performance played a mediator role.

The regression results are summarized in Table~\ref{tab:regression_midterm3}. We adopted a stepwise approach, incrementally adding predictors in four models to better understand the contributions of each factor.

In \textbf{Model 1}, we included \textbf{Proof-Review-AI-Tutor Usage Frequency} and \textbf{Chatbot Usage Frequency} as predictors. The results showed that \textbf{Chatbot Usage Frequency} significantly negatively predicted \textbf{Midterm 3 scores} (\( B = -0.056, p = 0.000 \)), while \textbf{Proof-Review-AI-Tutor Usage Frequency} had no significant effect (\( p = 0.713 \)). This model explained \textbf{13.4\%} of the variance in the scores (\( R^2 = 0.134 \)).

In \textbf{Model 2}, we introduced \textbf{Self-Efficacy} as an additional predictor. The results demonstrated that self-efficacy was a strong positive predictor of \textbf{Midterm 3 scores} (\( B = 5.088, p < 0.001 \)), while the negative effect of \textbf{Chatbot Usage Frequency} persisted (\( B = -0.040, p = 0.002 \)). This expanded model explained \textbf{26.1\%} of the variance (\( R^2 = 0.261 \)).

In \textbf{Model 3}, we further included \textbf{Homework Average Score}. \textbf{Homework Average Score} emerged as a significant positive predictor (\( B = 0.327, p = 0.002 \)), indicating that performance on homework assignments strongly predicted \textbf{Midterm 3 scores}. The effect of \textbf{Chatbot Usage Frequency} remained significant (\( B = -0.037, p = 0.003 \)), and self-efficacy continued to play a strong positive role (\( B = 4.513, p < 0.001 \)). This model explained \textbf{31.0\%} of the variance (\( R^2 = 0.310 \)).

Finally, in \textbf{Model 4}, we included students' previous \textbf{Midterm 1} and \textbf{Midterm 2 scores}. Both \textbf{Midterm 2} (\( B = 0.288, p = 0.022 \)) and \textbf{Midterm 1} (\( B = 0.150, p = 0.216 \)) were positive predictors, though only \textbf{Midterm 2} reached statistical significance. Additionally, \textbf{Homework Average Score} (\( B = 0.179, p = 0.111 \)) and \textbf{Self-Efficacy} (\( B = 3.207, p = 0.004 \)) remained significant. Notably, the negative effect of \textbf{Chatbot Usage Frequency} persisted (\( B = -0.030, p = 0.014 \)), reinforcing its association with lower \textbf{Midterm 3 outcomes}. This final model explained \textbf{35.5\%} of the variance (\( R^2 = 0.355 \)).

These findings suggest a multifaceted relationship between LLM-Tutor usage, self-efficacy, and exam performance. Although prior mid-term scores, self-efficacy and homework performance emerged as strong positive predictors of \textbf{Midterm 3 scores}, \textbf{Chatbot Usage Frequency} consistently demonstrated a significant negative association with exam outcomes. This pattern may indicate that reliance on chatbot features could reflect underlying struggles or ineffective learning strategies that require further investigation.

\subsection{Mediation Analysis on Self-Efficacy, Proof-Review-AI-Tutor Usage, Chatbot Usage, and Midterm Performance}

A mediation analysis using SmartPLS was performed to examine how Self-Efficacy influences Midterm 3 scores through Chatbot Usage Frequency and Proof-Review-AI-Tutor Usage Frequency.

\begin{figure}[h]
    \centering
    \includegraphics[width=0.95\linewidth]{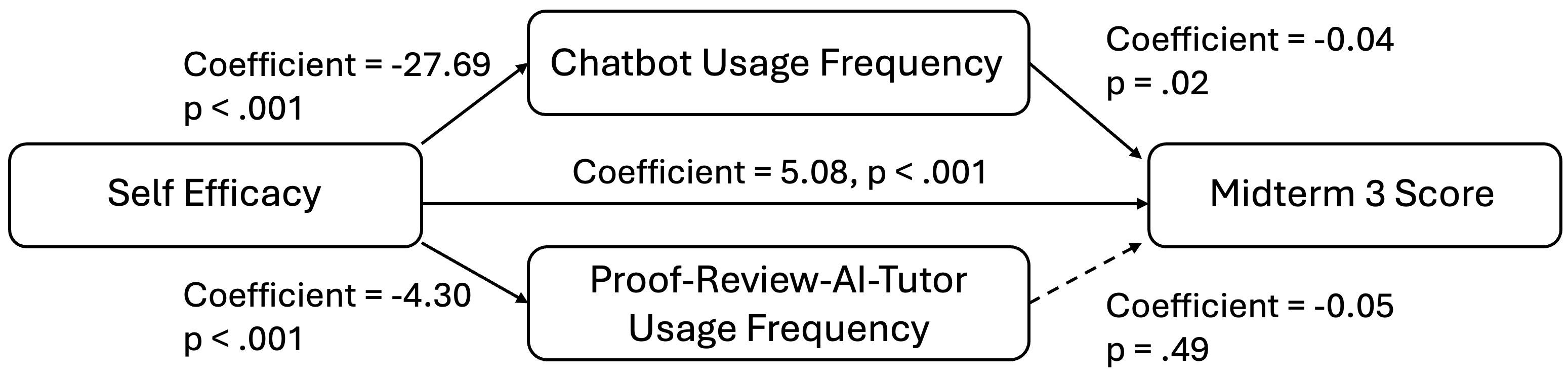}
    \caption{The image explains the relationship between Self-Efficacy and Midterm 3, with Proof-Review-AI-Tutor Usage and Chatbot Usage acting as mediators. Solid lines indicate significance (p < .05). This result indicated that the Chatbot Usage Frequency is partially mediated the effect of Self-Efficacy on Midterm 3 Score. Students with lower Self-Efficacy use the chatbot more, and this will results in a lower Midterm 3 exam performance.}
    \label{fig:self_efficacy_midterm3_mediation}
\end{figure}

The result of mediation is presented in Figure \ref{fig:self_efficacy_midterm3_mediation}. These findings suggest that \textbf{Self-Efficacy} influences \textbf{Midterm 3} scores both directly and indirectly. \textbf{Chatbot Usage Frequency} partially mediates this relationship, with higher reliance on chatbots because of lower \textbf{Self-Efficacy} negatively impacting \textbf{Midterm 3} scores. \textbf{Proof-Review-AI-Tutor Usage Frequency}, however, showed no significant mediating effect.

\begin{figure}[h]
    \centering
    \includegraphics[width=0.95\linewidth]{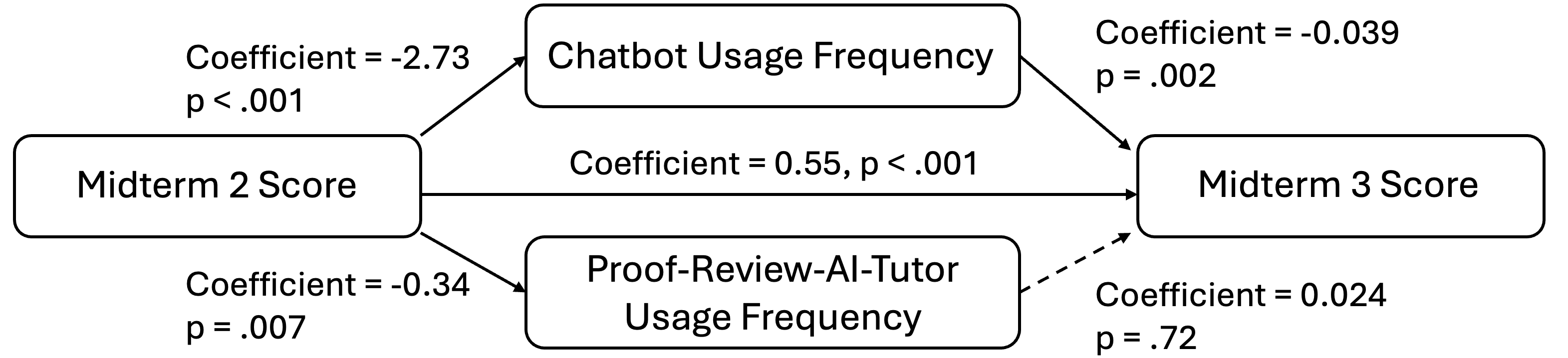}
    \caption{The image explains the relationship between Midterm 2 and Midterm 3, with Proof-Review-AI-Tutor Usage and Chatbot Usage acting as mediators. Solid lines indicate significance (p < .05).  This result indicated that the
    Chatbot Usage Frequency is partially mediated the effect of Midterm 2 to predict Midterm 3 Score. Students with lower Midterm 2 use the chatbot more, and this will results in a lower Midterm 3 exam performance.}
    \label{fig:midterm2_midterm3_mediation}
\end{figure}

We replaced \textbf{Self-Efficacy} with the \textbf{Midterm 2} scores and obtained the same patterns, as shown in Figure \ref{fig:midterm2_midterm3_mediation}.

\begin{figure}[h!]
    \centering
    \includegraphics[width=0.95\linewidth]{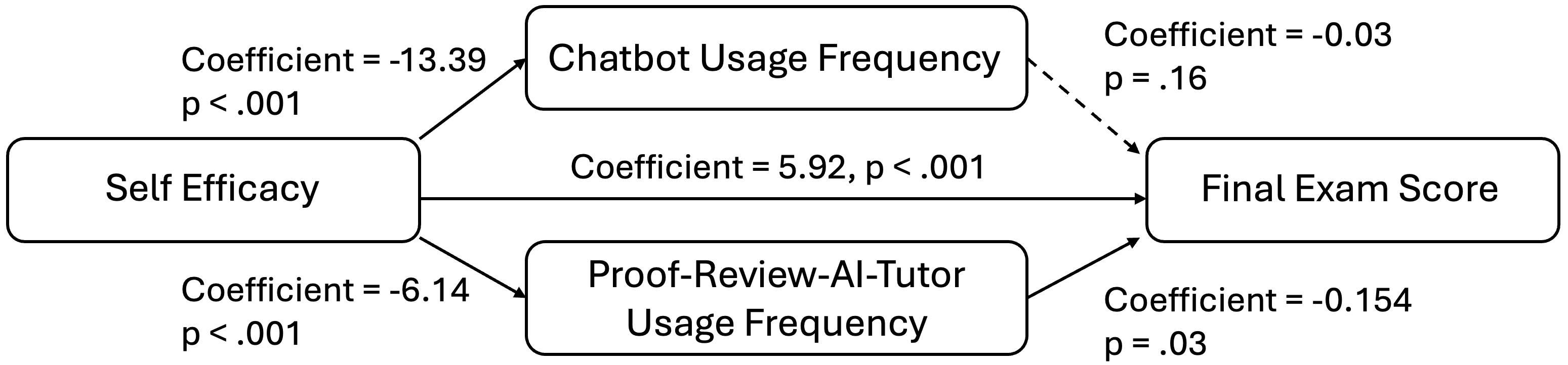}
    \caption{The mediation analysis results show the relationship between Self-Efficacy and Final Exam scores, with Chatbot Usage and Proof-Review-AI-Tutor Usage as mediators. Solid lines indicate significant paths (p < .05). The analysis highlights that while Chatbot Usage Frequency no longer mediates the effect, Proof-Review-AI-Tutor Usage Frequency exhibits a partial mediation effect. This result indicated that Students with lower Self-Efficacy use the Proof-Review-AI-Tutor more after midterm 3, and this will results in a higher Final Exam performance.}
    \label{fig:self_efficacy_midterm4_mediation}
\end{figure}

However, during the \textbf{Final Exam}, we observed a different trend. As shown in Figure \ref{fig:self_efficacy_midterm4_mediation}, in the Final Exam, \textbf{Self-Efficacy} still influences \textbf{Midterm 3} scores both directly and indirectly. However, the indirect effect of \textbf{Chatbot Usage Frequency} was no longer significant. Instead, \textbf{Proof-Review-AI-Tutor Usage Frequency} demonstrated a significant partially negative mediation effect. This suggests that, after midterm 3, students with lower \textbf{Self-Efficacy} tended to use \textbf{Proof-Review-AI-Tutors} more frequently, which partially explains their better performance in the \textbf{Final Exam}.

\subsection{Survey Results}

\subsubsection{Time Spent Comparison}
At the start of the experiment after midterm 1, participants were asked to report the amount of time they spent on assignments and textbook reading. After the midterm 2, participants were invited to report these data again. Data were analyzed only for participants who submitted responses at both time points. In the experimental group, 41 of 73 participants provided responses, while in the control group, 40 of 75 participants responded. 

\textbf{The results did not show significant differences in the time spent between the two groups}, either before using LLM-Tutor (Control Group: mean = 11.81 hours, std = 3.98; Experimental Group: mean = 12.07 hours, std = 4.71; t = 0.26, p = 0.79) or after the experimental group used LLM-Tutor (Control Group: mean = 7.55 hours, std = 2.85; Experimental Group: mean = 7.68 hours, std = 3.06; t = 0.20, p = 0.83).

\subsubsection{LLM-Tutor System Usability Score}

After the second midterm exam, we invited the experimental group to complete the SUS (System Usability Scale) questionnaire \citep{brooke1996sus} for LLM-Tutor. Among 73 experimental group participants, a total of 41 responses were collected, yielding a Cronbach's alpha of 0.875 and an \textbf{average SUS score of 69.94}. The SUS score ranges from 0 to 100; while it is not a percentage, higher scores indicate better usability. According to \citep{bangor2008empirical}, the average SUS score for various web applications is 68.05. On this scale, scores from 0-50 indicate poor usability, 51-70 reflect acceptable usability, 71-85 indicate good usability, and 86-100 represent excellent usability. Thus, the LLM-Tutor provides a slightly above-average user experience but still has room for improvement.

\subsection{Interview Results}

We invited a total of 19 participants from the experimental group to be interviewed about their experiences with LLM-Tutor. Subsequently, two undergraduate research assistants performed qualitative coding of the interview transcripts. They independently coded four interviews, achieving an inter-rater reliability Kappa score of 0.80, which indicates strong agreement \citep{mchugh2012interrater}.

\subsubsection{Changes in Students' Homework Habits with LLM-Tutor}

Regarding the research question of whether LLM-Tutor has brought about behavioral changes in students’ homework completion processes, a majority of students reported no significant changes (N = 11). Many maintained their pre-LLM-Tutor habits, such as independently attempting problems and seeking clarification from peers or during office hours.

\begin{quote}
``I don’t see many differences in my study methods. I just mostly see it as a resource to confirm that what I’m already doing is right.'' (P2)
\end{quote}

\begin{quote}
``Yeah, again, I wouldn't say it changes so much... Now I am using like a combination of LLM-Tutor and the other.'' (P8)
\end{quote}

On the other hand, a notable portion of students (N = 7) highlighted specific benefits that LLM-Tutor introduced to their workflow. For instance, students indicated that LLM-Tutor is very accessible.

\begin{quote}
    ``LLM-Tutor is that it's always available, and, like LLM-Tutor doesn't sleep. So if I need help, really late at night. I don't have to wait until the next day. I could just ask it'' (P1)
\end{quote}

\begin{quote}``I would say,[LLM-Tutor is] probably better [than office hour], because it's like continual. And it's also like more individualized. So, whereas the TA ... you ask them one question, and then you have to wait a long time to ask again. LLM-Tutor, you can ask a bunch of things like as you work through the problem.'' (P18)\end{quote}

Moreover, some students expressed feeling less hopeless or saving times when approaching challenging problems. This reassurance reduced the stress and frustration often associated with difficult assignments.

\begin{quote}
    ``With LLM-Tutor, I feel less hopeless and more willing to like start a problem, because I know that when I get stuck I can just ask it'' (P17)
\end{quote}

\begin{quote}
    ``Before using this, it took too long, like 3 to 4 hours, just to like get a sense of what I should do. And that decreased a lot, and that also decreased a lot of my stress, which just impacts positively for kind of everything.'' (P18)
\end{quote}


A noteworthy observation is that a few students (N = 4) described LLM-Tutor as a ``sanity check,'' likening its role to that of a grammar checker or peer reviewer. Although this step did not substantially alter their problem-solving approach, it added an extra layer of confidence to their process.

\begin{quote}
    ``I found that it was useful in pointing out my okay minor issues.'' (P11)
\end{quote}

This nuanced feedback highlights the diversity in how students integrated LLM-Tutor into their homework routines, with some incorporating it as a core component of their process and others using it as a supplementary resource.

\subsubsection{Learning Impact by LLM-Tutor}

Regarding student feedback on LLM-Tutor’s impact on their understanding of the course content, the majority of students (N = 13) reported an improvement in their comprehension of complex concepts and problem-solving strategies.

\begin{quote}``...I had a really hard times grasping what it was [the question]. So I just put the question, into LLM-Tutor, and I just asked it to break it down for me. ...So I really like the fact that, you know it would never get impatient with me. ... But I guess with LLM-Tutor that I could just really make it sit down with me and just go through the problem like literally, line by line, like, what is happening in this? Why are we doing this? Which theorem we're referring to...?''(P14)\end{quote}

\begin{quote}
    ``There was a [question]... it was just a little confusing... so I just went to LLM-Tutor and asked, 'Can you explain this method? How do the summations work? How does the binomial equation work?' and it explained it really well, and I was able to apply it in my homework. So it helps a lot.'' (P4)
\end{quote}

Also, students appreciated LLM-Tutor’s ability to clarify the nuances of proof techniques, including induction and partitioning, ensuring that their work met the required criteria.

\begin{quote}
    ``One time was when I did a proof by induction, and it told me that, like my induction hypothesis was unclear, so I fixed it.'' (P9)
\end{quote}

Furthermore, LLM-Tutor’s capability to deliver instant feedback and corrections enabled students to iteratively refine their understanding, which they found particularly valuable when addressing complex problems.

\begin{quote}
    ``[LLM-Tutor] explain to me in different words until I finally got what was happening in the problem and why the proof looks like it does and why it works.'' (P13)
\end{quote}

\begin{quote}
    ``I don't have to ask the teachers, or the TAs, about every little thing'' (P3)
\end{quote}

Conversely, a smaller group of students (N = 5) reported that LLM-Tutor did not make a noticeable difference in their understanding. These participants reported relying more on traditional resources, such as lecture notes and TA office hours, to deepen their understanding.


\begin{quote}
    ``I don't know if it's had that much of an impact on my understanding of the course content as a whole.'' (P19)
\end{quote}

\subsubsection{Confidence Impact by LLM-Tutor}

To assess the behavioral impact of LLM-Tutor on students’ confidence in mathematical problem-solving tasks, the majority of students (N = 11) reported that LLM-Tutor did not significantly affect their confidence in their math skills. Although these students acknowledged the value of the tool in verifying solutions and identifying errors, they emphasized that it functioned primarily as a supplementary resource rather than a confidence booster.

\begin{quote}
    ``No, it doesn't really change my confidence in my math abilities.'' (P1)
\end{quote}
\begin{quote}
    ``... maybe [it made me] more confident in the solution that I submitted. I'm not sure if [it] made me more confident in my personal [abilities].'' (P6)
\end{quote}

On the other hand, a notable proportion of students (N = 7) experienced a slight increase in confidence. They highlighted several ways in which LLM-Tutor positively influenced their mindset, such as feeling more prepared for future tests due to the iterative feedback provided, reducing anxiety when starting homework sets, and gaining a stronger sense of confidence after verifying their solutions.

\begin{quote}``... it helped me like realize that ... all I need is maybe a little bit like one more prompting step to get me to get me to the like the finish line.'' (P17)
\end{quote}

\subsubsection{Comparison of LLM-Tutor Feedback with TA Office Hour}


Student opinions on the quality of the feedback provided by LLM-Tutor compared to those of TAs revealed both strengths and limitations of the tool. Many students (N = 12) found LLM-Tutor's feedback lacking depth and contextual alignment with course expectations and the knowledge point behind the mistake.

\begin{quote}``I think the TA's feedback is still a bit more comprehensive and more targeted towards the ... theoretical misconception behind the mistake. whereas LLM-Tutor, it just identifies the mistake'' (P12)\end{quote}

Several students (N = 5) described the feedback from LLM-Tutor as complementary to, rather than substitute for, the TA guidance. 
For example, they appreciated LLM-Tutor’s role in addressing simpler issues or serving as an initial review tool, but preferred TAs for more complex problem discussions.

\begin{quote}
    ``I think I like using LLM-Tutor more when I have an idea of what I need to do... if I'm facing a problem which I don't even know how to approach... I would then go and talk to TA'' (P8)
\end{quote}

Interestingly, a minority (N = 2) considered the LLM-Tutor feedback superior to TA in specific contexts, such as clarity and accessibility. These students noted that LLM-Tutor provided instant, structured feedback on specific proof components, which was sometimes more useful than generalized TA comments. 

\begin{quote}``I even think LLM-Tutor was more clear [on] feedback on my proof. Because [LLM-Tutor] like specifically mentioned this line, [and which] part is not clear. ... But in TA's comments, it's more like this is wrong in [a] vague sentence,... so I don't really understand what's wrong with [the proof].''(P16)\end{quote}

\subsubsection{Impact of LLM-Tutor on Motivation and Engagement in Mathematics Learning}

Students' experiences with LLM-Tutor revealed varying effects on their motivation and engagement with the course material. Many students (N = 12) reported that LLM-Tutor helped them engage more deeply with the course content. 

\begin{quote}
``I think it motivates me more to do my homework, and put effort into it. So yeah, [LLM-Tutor] helps.''(P16)
\end{quote}
\begin{quote}
``...after getting answers [from LLM-Tutor], I would look back at lecture notes.[And]
see what I could improve.''(P10)
\end{quote}

However, other students (N = 7) indicated that LLM-Tutor did not significantly affect their enthusiasm for mathematics. These students viewed it primarily as a resource for verifying answers or clarifying doubts rather than as an inspirational or transformative tool.

\begin{quote}
``I would say [motivation] it's fairly neutral. I think [LLM-Tutor is] mostly good when I'm really, really stuck.'' (P8)
\end{quote}

However, some students (N = 5) expressed concerns about their reliance on LLM-Tutor, especially the chatbot. They worried that relying too much on the tool could hinder their ability to develop independent problem-solving skills.

\begin{quote}
    ``I think because I have access to LLM-Tutor, I actually tend to forget a lot of things. LLM-Tutor [chatbot] makes it very easy to find information, so I don’t spend as much time trying to figure out where to start or how to solve the problem on my own.'' (P1)
\end{quote}

\begin{quote}
``Sometimes I feel like I need to make a conscious effort not to become reliant on it.
It's the kind of feeling that I get where you're scared your writing ability will deteriorate after using large language models to aid your writing. Same idea, but it's for math proofs.'' (P12)\end{quote}

\section{Discussion}

The results of our study highlight several important insights regarding the deployment and effectiveness of AI-powered tutoring systems like LLM-Tutor. While these systems offer impressive capabilities, their actual impact on learning outcomes remains nuanced and contingent on various factors. Below, we discuss the key findings and suggest future directions for improving such systems.

\textbf{Enhancing Students' Performance on Homework but not on Exam}. Our findings indicate that the rapid deployment of AI tutoring systems, such as LLM-Tutor, yields mixed results. While LLM-Tutor showed statistically significant improvements in homework assignment scores, its impact on broader learning outcomes, such as exam performance or time spent on homework, was not significant. This indicates that such systems may effectively support task-specific learning activities, like homework, but their broader influence on other learning outcomes (e.g., exams) remains limited. The findings suggest generative AI tools like ChatGPT alone may not be enough; incorporating learning sciences principles is essential to enhance these systems further and optimize their impact on learning outcomes.

\textbf{Overreliance on Chatbot-Based Support.} Our analysis revealed that frequent use of the chatbot feature was associated with lower exam performance, particularly among students with low self-efficacy. This suggests a potential over-reliance on chatbots, which may hinder independent problem-solving skills. However, it is unclear whether chatbot usage is a cause or a consequence of students' low self-efficacy. A preliminary examination of chat logs indicates that most students primarily use the chatbot for clarification of basic terminology or concepts or paste their homework directly, which may not effectively support the development of independent proof-writing and problem-solving skills. Future work should focus on refining chatbot design to better support students' learning needs and examining usage patterns to understand how these interactions impact learning outcomes.

\textbf{Enhancing Final Exam Performance for low self-efficacy students by LLM-Tutor Proof-Review-AI-Tutor Interface.} In contrast to the chatbot, the Proof-Review-AI-Tutor feature demonstrated a positive effect on students' final exam performance, particularly for those with lower self-efficacy. This suggests that interfaces like Proof-Review-AI-Tutor (Figure \ref{fig:homework_reviewer}) have the potential to be a targeted support tool for low self-efficacy students. By providing timely and specific feedback on homework, the system may help students build confidence and improve their understanding. Nevertheless, the effectiveness of the Proof-Review-AI-Tutor was not observed before Midterm 3, possibly due to shifts in students' learning strategies. Further research is needed to explore how students at different self-efficacy level adapt and regulated their learning habits with AI tools overtime.

\textbf{Reflections from Student Interviews.} Interviews with students revealed mixed perspectives on LLM-Tutor's impact. While many appreciated its ability to provide instant feedback and clarify challenging concepts, others expressed concerns about over-reliance and the system's limitations in addressing broader conceptual gaps. These findings suggest that tools like LLM-Tutor may be more suitable when used as a supplemental tool rather than a standalone solution.

\section{Future Works}

Our results also highlight the complexity of measuring student learning solely through exam grades. Learning is a multifaceted process, and students may adjust their strategies over time in response to tools like LLM-Tutor. This could potentially explain why we observed a shift in mediation effects in the final exam—where the chatbot's influence was no longer significant, while the effect of the Proof-Review-AI-Tutor became significant. To better understand how students interact with LLM-Tutor, future research should focus on qualitative analyses of interaction logs. By identifying patterns in how students use the system and how these patterns correlate with learning outcomes, researchers can develop more effective AI-driven tutoring strategies. Additionally, incorporating mixed-method approaches, such as interviews and surveys, could provide deeper insights into students' perceptions and experiences.

In addition, future work should explore a hybrid approach where students use LLM-Tutor for immediate feedback on simpler issues, such as clarifying concepts or fixing minor errors, before seeking help from human resources like TAs or peers for more complex or conceptual challenges. For instance, the system could include a feature that flags unresolved questions or repeated misunderstandings, prompting students to read relevant textbooks, attend office hours or consult instructors. Additionally, integrating analytics from LLM-Tutor interactions could help instructors identify common student struggles, enabling more targeted support during recitations or discussions sessions. This approach ensures that AI tools complement rather than replace traditional resources, reducing dependency and maintaining active engagement with human instruction.


\section{Conclusion}

This study evaluated the use of LLM-Tutor, an AI-powered tutoring system, in a discrete mathematics course, focusing on its impact on student performance, self-efficacy, and engagement. While LLM-Tutor demonstrated significant benefits in improving homework performance and providing timely feedback, its broader effects on exam outcomes and time spent were insignificant. Students appreciated its role as a supplemental resource but highlighted limitations in addressing conceptual gaps and concerns about over-reliance.

These findings underscore the need for careful design and integration of AI tools into educational contexts. Generative AI systems like LLM-Tutor have great potential to enhance learning but must be complemented by pedagogical principles and traditional resources to maximize their impact. Future research should focus on refining these tools to better support independent problem-solving and long-term learning outcomes.

\section*{Acknowledgements}
Our research was approved by the institutional review board.

\section*{Conflict of interest}
The authors declare no conflict of interest.

\section*{Ethics Statement}
Ethics approval was obtained prior to the analysis of the archived data.

\section*{Data Availability Statment}
Due to privacy issues and IRB constrains, the data cannot be made publicly available.


\printendnotes

\bibliography{main}

\end{document}